\documentclass[preprint,authoryear,12pt]{elsarticle}

\usepackage{graphicx}

\usepackage{amssymb}

\usepackage[usenames, dvipsnames]{color}

\usepackage{hyperref}

\journal{Advances in Space Research}

\begin{document}

\begin{frontmatter}




\title{Gaia: the Galaxy in six (and more) dimensions}


\author{Elena Pancino\corref{cor}}
\address{INAF -- Osservatorio Astrofisico di Arcetri, Largo Enrico Fermi 5,
50125 Firenze, Italy\\SSDC -- Space Science data Center, ASI, Via del
Politecnico snc, 00133 Roma, Italy\\Based on the invited interdisciplinary lecture {\em
``Gaia: the Galaxy in 6D+ dimensions''} given at the COSPAR 42$^{nd}$ assembly held in
Pasadena, California, USA, in July 2018.}
\cortext[cor]{Corresponding author}
\ead{elena.pancino@inaf.it}




\begin{abstract}

The ESA cornerstone mission {\em Gaia} was successfully launched in 2013, and
is now scanning the sky to accurately measure the positions and motions of
about two billion point-like sources of 3$\lesssim$V$\lesssim$20.5~mag, with
the main goal of reconstructing the 6D phase space structure of the Milky Way.
The typical uncertainties in the astrometry will be in the range
30-500~$\mu$as. The sky will be repeatedly scanned (70 times on average) for
five years or more, adding the time dimension, and the Gaia data are
complemented by mmag photometry in three broad bands, plus line-of-sight
velocities from medium resolution spectroscopy for brighter stars. This
impressive dataset is having a large impact on various areas of astrophysics,
from solar system objects to distant quasars, from nearby stars to unresolved
galaxies, from binaries and extrasolar planets to light bending experiments.
This invited review paper presents an overview of the {\em Gaia} mission and
describes why, to reach the goal performances in astrometry and to adequately
map the Milky Way kinematics, {\em Gaia} was also equipped with
state-of-the-art photometers and spectrographs, enabling us to explore much
more than the 6D phase-space of positions and velocities. Scientific highlights
of the first two {\em Gaia} data releases are briefly presented.

\end{abstract}


\end{frontmatter}

\parindent=0.5 cm

\section{Introduction}
\label{sec:intro}

Astrometry, one of the oldest branches of astronomy, is the science of measuring
the positions and motions of objects on the celestial sphere. Besides providing
the apparent motion of celestial objects, which is one of the ingredients for
stellar kinematics and dynamics, astrometry answers two fundamental questions of
astrophysics. The first concerns the definition of a reference system of
celestial positions -- astronomical coordinates are the most powerful tool to
identify individual objects and to combine and compare large catalogues coming
from observational surveys. The ICRF \citep[International Coordinates Reference
Frame,][]{arias95,vlbi} is the standard in the field, based on very long
baseline interferometry (VLBI) of compact radio sources. The second ingredient
is distance, an elusive but fundamental ingredient, without which it would be
impossible to fully understand the universe. Astrometry, by measuring parallax
-- the reflection of the Earth's motion around the Sun -- is one of the few
available techniques to provide direct estimates of distances, without
assumptions on objects properties such as intrinsic luminosity. The more
accurate and precise the astrometric measurement, the smaller the parallax that
can be accurately measured, the larger the distance that can be reached.

\begin{figure}
\centering
\includegraphics[width=9.4cm]{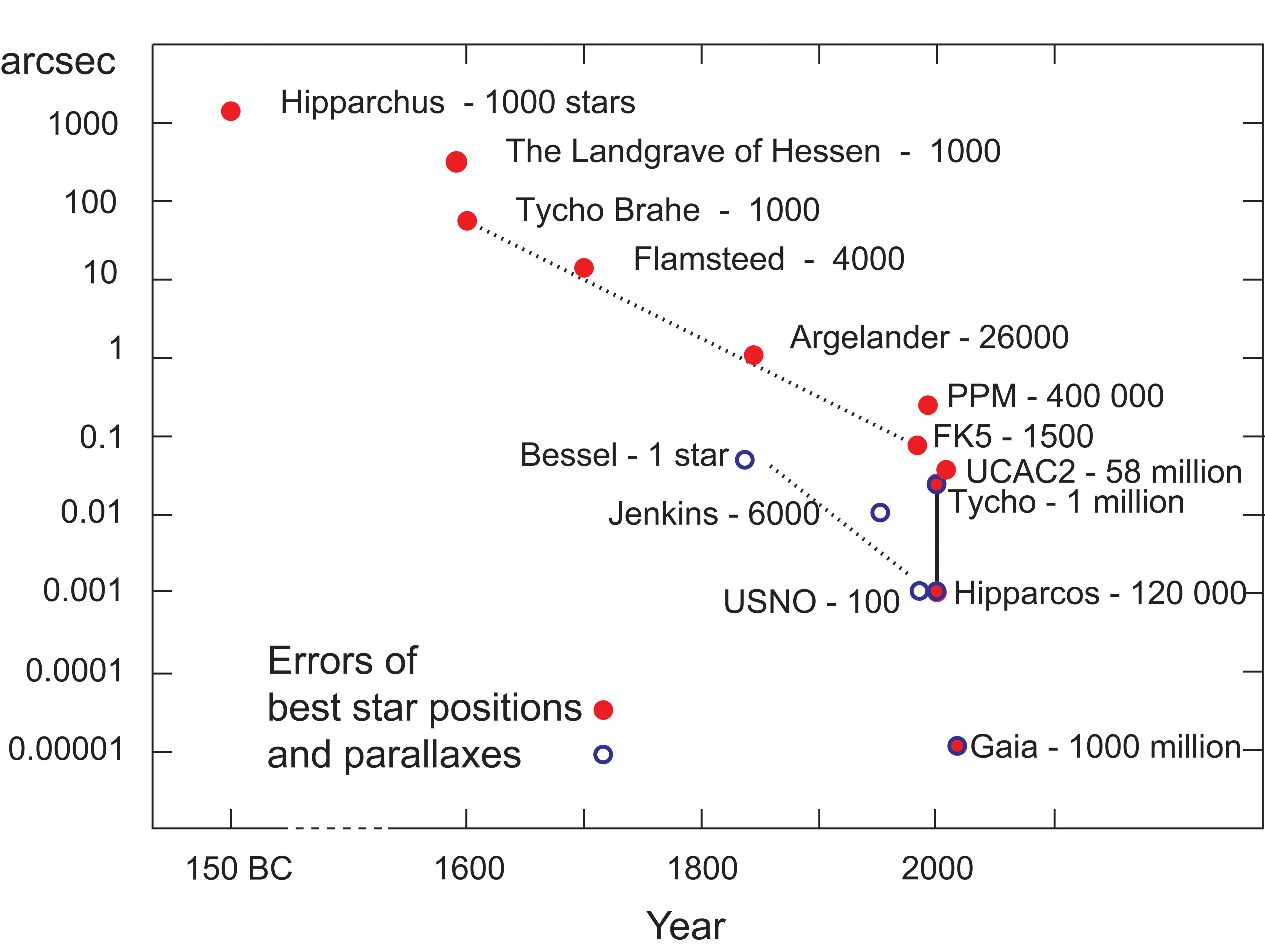}
\caption{The improvements of astrometry measurements in the course of the
centuries. Dotted lines show the rate of improvement for positions (red dots)
and parallaxes (blue circles), while the vertical line shows the enormously
accelerated improvement obtained in the last century. {\em Gaia}'s contribution
to the field brings a further improvement of a few orders of magnitude compared
to Hipparcos (see text). Image source: ESA.}
\label{fig:astrom}
\end{figure}

{\em Gaia} \citep{gaia}, is the ESA cornerstone astrometric mission, launched in
2013, whose main goal is to provide absolute astrometry, 100 times more accurate
than its extremely successful predecessor \citep[Hipparcos,][]{hip} and to
target much fainter objects, down to a magnitude of $V\simeq$~20.5~mag, thus
providing a catalogue of two billion objects, covering the entire sky. The
ambition of {\em Gaia} is to provide a homogeneous census of the phase space of
positions and motions of as much as 1\% of the stars in the Milky Way. {\em
Gaia} is also equipped with a spectrograph, to measure line-of-sight velocities
of millions of stars of various spectral types, down to $G
\simeq$~16-17~mag\footnote{The {\em Gaia} white-light magnitude is not too
different from the Johnson V magnitude for stars with non-extreme colors. A full
set of transformations between the {\em Gaia} photometry and widely used
photometric systems can be found in \citet{evans18}. The {\em Gaia} photometry
is described more in detail in Section~\ref{sec:photo}.}. To give an idea of the
improvement that {\em Gaia} is meant to provide, Figure~\ref{fig:astrom} shows
how astrometric measurement uncertainties have evolved through time: {\em
Gaia}'s predecessor, Hipparcos, could obtain positions and parallaxes with
uncertainties of the order of 1~mas, for 10$^5$ stars down to V$\simeq$12~mag.
{\em Gaia}, on the other hand, will provide astrometry with uncertainties of the
order of 0.01~mas, for 10$^9$ stars down to V$\simeq$21~mag (at the end of the
mission).

To be able to provide such high quality astrometry, however, {\em Gaia} will
need also to collect exquisite photometry in three wide bands, which is
necessary to correct for chromatic effects on the measurements of object's
positions. {\em Gaia} will also provide time sampling: on average each object
will be observed 70 times over the nominal mission duration of 5 years, to allow
for accurate global astrometry and to break the degeneracy between proper motion
and parallax. In substance, {\em Gaia}'s exquisite astrometry {\em requires}
excellent photometry and spectroscopy, and the full catalogue will be
unprecedented in optical astronomy in terms of the number of measured sources,
characterized by (quasi-)simultaneous measurements with several different
techniques. This is one of the reasons why {\em Gaia} is bringing a real
revolution in many fields of modern astrophysics, from stellar structure and
evolution, to the reconstruction of the Milky Way history and present day
chemo-dynamical status; from the study of the motion of Solar System object, to
a wealth of data on distant, unresolved galaxies and QSO; from its
planet-hunting capabilities, to the experiments of fundamental physics that it
will enable. All {\em Gaia} released data can be obtained from various sources,
like the ESA {\em Gaia} archive or the {\em Gaia} Partner Data
Centers\footnote{\url{https://www.cosmos.esa.int/web/gaia/data-access}}.

In this paper, I briefly describe the mission capabilities and performances, and
I provide a few examples of the exciting results that have been obtained by the
community using {\em Gaia} data coming from its first \citep[DR1,][]{gaia1} and
second \citep[DR2,][]{gaia2} data releases. The paper is organized as follows:
Section~\ref{sec:gaia} presents the spacecraft, onboard instrumentation, and
science performances; Section~\ref{sec:results} presents the {\em Gaia} data
releases and their content, and highlights some early results obtained with the
first two releases; Section~\ref{sec:concl} summarizes the main points and
concludes with a future outlook.

\section{{\em Gaia}}
\label{sec:gaia}

The idea for an astrometric mission that could improve on Hipparcos dates back
to the 80s, and started appearing in international Journals of the field in 1994
\citep{oldgaia}, with an interferometric design that later was abandoned. The
mission was formally approved by ESA in 2000, as part of its Horizon 2000 Plus
program. Its main scientific goal is to study the structure, formation and
evolution of our galaxy, the Milky way, by studying stars belonging to all its
components and to the Local Group \citep{perryman01}. In 2006, after the
Announcement of Opportunity by ESA, the {\em Gaia} Data Processing and Analysis
Consortium (DPAC) was formed, to take care of all aspects of data treatment and
to deliver the data to the community. At about the same time, the contract for
the construction of {\em Gaia} was granted to Astrium (now Airbus Defence and
Space). Launch occurred successfully in December 2013, and the satellite reached
very smoothly its Lissajus orbit around L2, the unstable Lagrangian point of the
Sun and Earth-Moon system, where it started its technical and scientific
commissioning phase, and started nominal operations about six months later. 

\begin{figure}
\centering
\includegraphics[height=5.8cm]{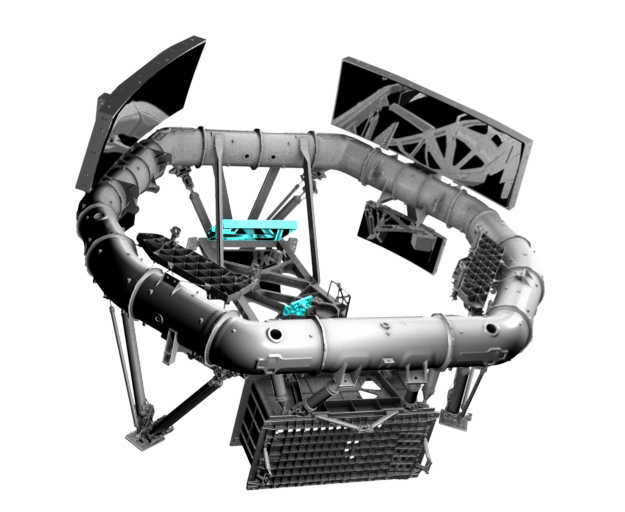}
\includegraphics[height=5.8cm]{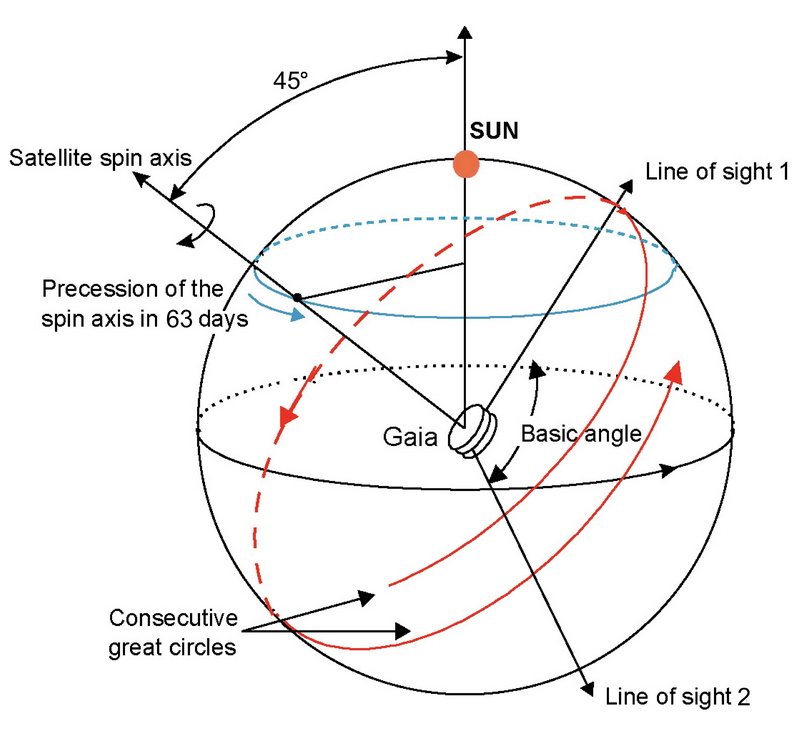}
\caption{Left panel: Gaia payload. The two Gaia mirrors are mounted on the
paylod torus (top), observing two different lines of sight that are focalized
through a series of mirrors (M4 and M5 are visbile, in cyan) onto a single focal
plane (its back side is visible at the bottom of the figure, see also
Figure~\ref{fig:focal}). Right panel: {\em Gaia} scanning the sky, by precessing
its spin axis around the Sun direction. Images source: ESA.}
\label{fig:gaia}
\end{figure}

The nominal five years of {\em Gaia} observations were
completed in July 2019, but the onboard fuel reserve is expected to keep Gaia
operational until 2024. After a careful cost-benefit analysis, ESA formally
approved the first mission extension of two years, on November 2018. This
ensures that {\em Gaia} will receive support to operate until at least 2020.
Contextually, a preliminary pre-approval of a further two-year extension was
given, to be re-examined later, to keep on observing until 2022. The Gaia
mission extension is expected to improve positions, parallaxes, photometry, and
radial velocities by about 40\% and to improve upon proper motions even more.
The mission extension will also have a positive effect on the detection of
exoplanets and asteroids as well as on the time-sampling of variable
objects\footnote{More information on {\em Gaia} can be found in the official ESA
webpages for the public
(\url{http://www.esa.int/Our_Activities/Space_Science/Gaia})  and for scientists
(\url{https://www.cosmos.esa.int/web/gaia/}).}. 

\subsection{Astrometry}

Like Hipparcos, {\em Gaia} is one of the few missions designed to perform global
astrometric measurements. To this aim, it is equipped with two telescopes
observing two different lines of sight, separated by a wide angle -- the basic
angle of 106.5~deg -- and whose light is then combined on a single focal plane
(Figure~\ref{fig:gaia}, left panel) to allow for the simultaneous measurements
of large angular differences between celestial objects. The satellite scans the
whole sky by slowly precessing its spin axis (Figure~\ref{fig:gaia}, right
panel) and by doing so, it passes repeatedly on the same regions of the sky (on
average 70 times) each time with a different orientation. The large basic angle,
the repeated measurements, and the onboard metrology system \citep{mignard11},
provide global astrometric measurements and good control over the systematic
uncertainties of the astrometry. Thanks to its design, {\em Gaia} can thus
define its own kinematically non-rotating reference system
\citep{gaia_icrf,gaia_refsys}, based on more than half a million Quasars, that
are in part also observed with VLBI, thus allowing the axes of the reference
system to be aligned with the ICRF \citep{vlbi}. 

\begin{figure}
\centering
\includegraphics[height=6cm]{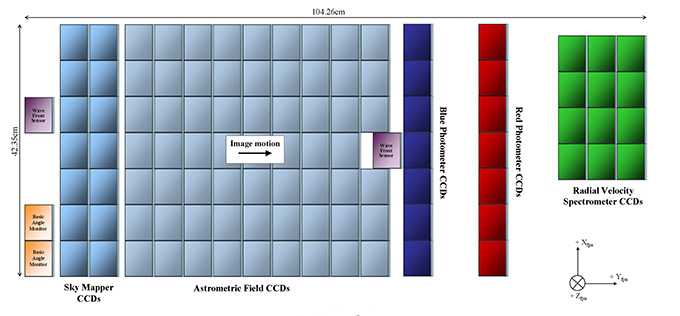}
\caption{The Gaia focal plane. Each rectangle represents one CCD. Light passing
through different instruments is collected by groups of CCDs marked in different
colors, as annotated on the Figure. Thanks to {\em Gaia}'s continuous sky
scanning, celestial objects appear on the focal plane on the left side and
travel horizontally across it to exit on the right side. Image source: ESA,
Alexander Short.}
\label{fig:focal}
\end{figure}

As described more in details by \citet{gaia}, the light from the two telescopes
reaches the focal plane, where 106 CCDs (Charge-Coupled Devices) gather the
light and are read out continously as {\em Gaia} scans the sky. From the point
of view of the focal plane, thus, it looks as if celestial objects enter the
focal plane on one side, travel across it at a speed that carefully matches the
read-out speed (about 4~s per CCD), and then exit on the other side. To save
telemetry bandwidth, objects are detected on board in the first two CCD columns
(Sky Mapper CCDs, or SM, colored in light blue in Figure~\ref{fig:focal}), and
confirmed in the third column: only the charges accumulated in little windows
that follow the object on the focal plane are transmitted to the ground. The
next set of 62 CCDs, the Astrometric Field (AF, colored in grey in
Figure~\ref{fig:focal}) collects photons in white light (300-1100~nm), to allow
the PSF (Point Spread Function) modeling and accurate centroiding, that is the
basis for the astrometric solution\footnote{see also
\url{https://www.cosmos.esa.int/web/gaia/astrometric-instrument}}.

The complex data processing is described by \citet{lindegren16,lindegren18}. 
Its heart is the Astrometric Global Iterative Solution (AGIS). Each time a new
chunk of data comes in from the satellite and is preprocessed, AGIS is re-run on
all the data available, using its previous solution as a starting point for the
new solution. This happens approximately every 6 months. AGIS solves for the
satellite attitude and for all five astrometric parameters: on sky positions
($\alpha$ and $\delta$), proper motions ($\mu_{\alpha}$ and $\mu_{\delta}$) and
parallax ($\varpi$). To help pinpointing {\em Gaia}'s position in the sky, a
network of medium-sized telescopes regularly observes it from Earth \citep[the
GBOT, Ground-Based Optical Tracking project,][]{gbot,buzz}. This is a novel
approach for ESA: {\em Gaia} is the first mission that is tracked from the
ground not only with radio tracking and ranging, but also with optical imaging. 
The volume of AF data processed is of tens of billions of measurements, and
every AGIS cycle contains of course more data than the previous one, increasing
the computational challenge. The satellite's attitude and Basic Angle monitoring
on board and in post-processing from the ground is so accurate and sensitive
that thermal changes and micro-meteoroid impacts in the L2 region produce jumps,
oscillations, and the so-called micro-clancks, very well visible in the
astrometric solution. Thus, they provide an entirely new insight into the space
weather conditions at L2.

The expected end-of-mission performances of {\em Gaia} are presented and updated
on the ESA
webpages\footnote{\url{https://www.cosmos.esa.int/web/gaia/science-performance}}.
With DR2, that is based on roughly two years of data over the nominal five, {\em
Gaia} is already approaching the exquisite quality expected at the faint end.
This is because the mission is not optimized for bright stars
(V$\lesssim$13\,mag), and more work and more data are required to improve the
bright end. In spite of the preliminary nature of the released data, DR2 is
already spawned numerous new discoveries in the first few weeks
(Section~\ref{sec:results}).

\subsection{Photometry}
\label{sec:photo}

To obtain the required quality in astrometric measurements, it is necessary to
map extremely well the differential object displacement that occurs in different
regions of the large {\em Gaia} focal plane, depending on the object intrinsic
color. In other words, the blue light and the red light emitted by each
celestial source travel through slightly different optical paths and end on
sligthly different positions on the focal plane. The net result is that cool or
reddened stars and galaxies have systematically different centroid measurements
than hot and blue objects. 

\begin{figure}
\centering
\includegraphics[height=7.5cm]{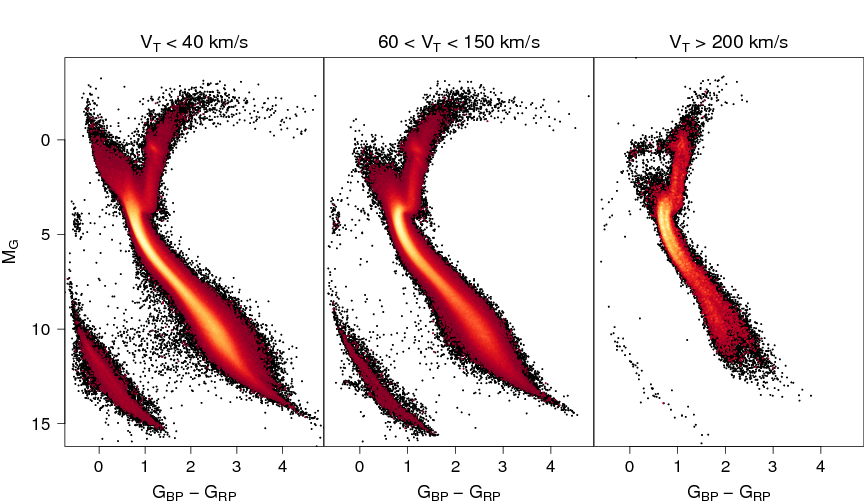}
\caption{Gaia Hertzsprung-Russell diagrams, Gaia absolute magnitude M$_G$ versus
$G_{\rm{BP}}$-$G_{\rm{RP}}$ color, as a function of the stars tangential
velocity (V$_T$), using Gaia DR2 with relative parallax uncertainty better than
10\% and low extinction stars (E(B-V)$<$0.015~mag), together with astrometric
and photometric quality filters. The color scale represents the square root of
the density of stars. Image source: ESA/Gaia/DPAC, Carine Babusiaux.}
\label{fig:cmd}
\end{figure}

With {\em Gaia}, we aim at 10-500~$\mu$as uncertainties in the absolute
astrometry, and the chromaticity effects cannot be neglected. This is one of the
reasons why -- besides the white light magnitudes measured on the AF -- {\em
Gaia} is equipped with two low-dispersion spectrographs: the blue and the red
(spectro-)photometers (hereafter, BP and RP). They produce spectra with a
resolution (R=$\lambda/\delta\lambda$) varying between 20 and 100, each covering
roughly half of the optical range covered by {\em Gaia} SM and AF
images\footnote{see \url{https://www.cosmos.esa.int/web/gaia/iow_20180316}}, the
BP from 330 to 670~nm and the RP from 620 to 1050~nm. 

The BP-RP color is obtained by extracting integrated magnitudes\footnote{The
term {\em integrated magnitude}, widely used in the {\em Gaia} community, simply
means that all photons collected in the white AF passband or in the BP/RP low
resolution spectra, or in the RVS spectrograph, are summed and converted to a
magnitude scale. For each instrument, respectively, the obtained integrated
magnitudes are indicated as G, G$_{\rm{BP}}$, G$_{\rm{RP}}$, and
G$_{\rm{RVS}}$.} from the BP and RP spectra. The obtained three-band photometry
of white light, integrated BP and RP magnitudes already in GDR2 contains 1.3
billion sources, down to $G\simeq~21$~mag and has internal uncertainties of the
order of a few millimagnitudes. This constitutes a problem because none of the
exisiting photometric catalogues has comparable uncertainties, and thus one
cannot efficiently use external catalogues to fully validate {\em Gaia}
photometry \citep{evans18}. The external (absolute) flux calibration is based on
an extension of the CALSPEC set of spectro-photometric standard stars
\citep{bohlin14,pancino12}, that provides external uncertainties of about 1\% or
better in flux, i.e., the best that can be done with current technology. The
three-band color-magnitude diagram from the latest {\em Gaia} release
(Figure~\ref{fig:cmd}) shows an impressive amount of detail, and has spawned new
discoveries even in the relatively old field of stellar structure and evolution
(see section~\ref{sec:results}). 

Besides integrated magnitudes and colors, {\em Gaia} BP and RP spectra provide a
wealth of additional astrophysical information. They can be used -- together
with the medium-resolution spectra described in the next section -- to classify
astrophysical objects, for example separating stars from galaxies or quasars.
The spectra also provide accurate astrophysical parameters such as surface
temperatures and gravities, metallicities, or interstellar reddening and
extinction. A preliminary set of astrophysical parameters was provided with {\em
Gaia} DR2 \citep{andrae18}, using solely integrated magnitudes. We expect
significant improvement in the quantity and quality of the released
astrophysical parameters in the upcoming releases.

\subsection{Spectroscopy}

{\em Gaia} astrometry is designed to provide five of the six phase-space
dimensions -- the space of positions and motions -- for roughly 2 billion stars:
sky position (right ascension and declination, or $\alpha$ and $\delta$),
distance (through parallax, or {$\varpi$), and the on-sky motion (proper motion
in $\alpha$ and $\delta$, or $\mu_{\alpha}$ and $\mu_{\delta}$). It cannot
however provide information on the motion along the line of sight, also called
radial velocity (RV hereafter). For this reason, {\em Gaia} is equipped with a
medium resolution spectrograph that provides line-of-sight velocities based on
the Doppler shift: the Radial Velocity Spectrometer or RVS \citep{cropper18}. 

The spectra have a resolution R$\simeq$11\,500 and cover the widely used region
around the Calcium IR triplet (846--874~nm), which allows for accurate RV
measurements for late type stars. For early type stars, the Paschen lines are
well visible in this range and for very cool stars the molecular bands of TiO
can be used for RV determination, albeit with lower performances compared to the
Calcium triplet. The region is also rich in atomic lines belonging mostly to the
iron-peak and $\alpha$-elements, and can thus be used to accurately parametrize
stars and to derive their chemical composition. Some diffuse interstellar bands
are also present, for the independent determination of interstellar absorption. 

The science performances depend on the spectral type and the brightness of the
observed stars, where the expected limiting magnitude is brighter than the one
of AF, BP, and RP: $G\simeq$16~mag. For the brigthest stars
(G$\lesssim$13--14~mag) it is about 1~km/s or less, which is quite competitive
with existing RV
surveys\footnote{\url{https://www.cosmos.esa.int/web/gaia/science-performance}}.
For the less favorable cases of very blue and faint stars it can reach 15~km/s
or more. All RV measurements are accurately calibrated using a set of more than
1000 RV standard stars \citep{soubiran18}. The currently released RV
measurements constitute the largest available set of homogeneously and
accurately measured RV in the literature to date, with more than 7 million stars
in the DR2 catalogue.

\subsection{Time coverage and variability}

\begin{figure}
\centering
\includegraphics[height=6.5cm]{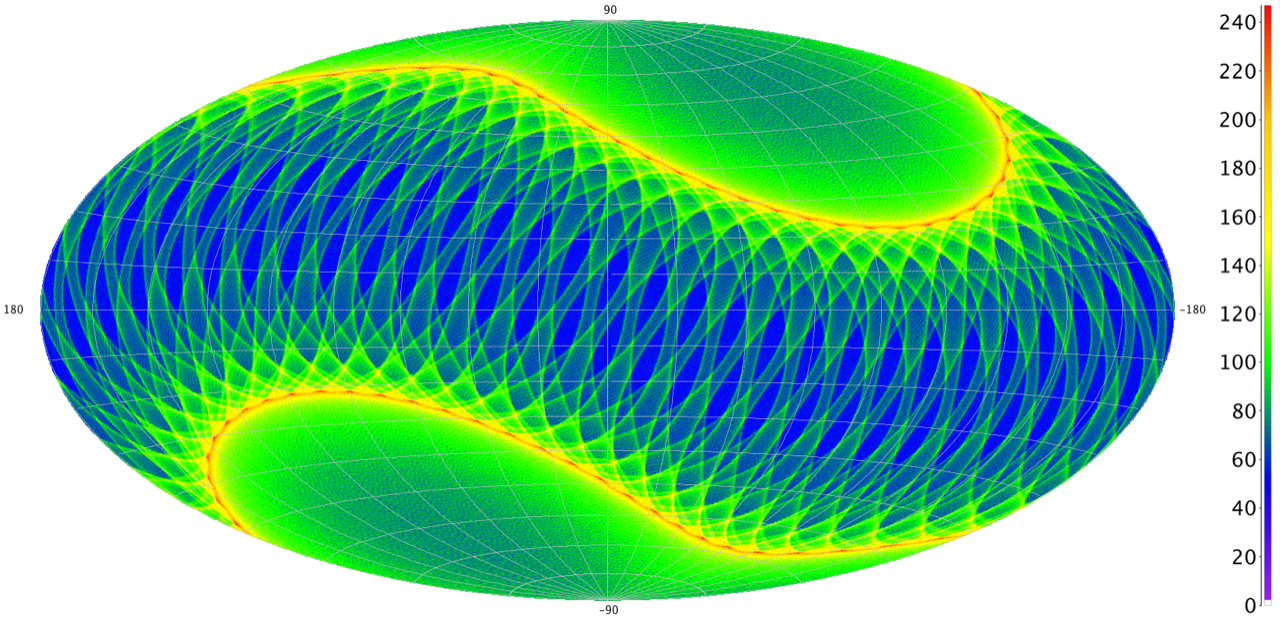}
\caption{Gaia field-of-view transits for the five-year Nominal Scanning Law in
ICRF coordinates. The Ecliptic plane is within the blue curved strip, while the
number of passages is maximum on the two lines at $\pm$45~deg from it. Image
source: Berry Holl.}
\label{fig:scan}
\end{figure}

The last necessary ingredient to obtain extremely accurate astrometry is time
coverage, i.e., repeated observations. One of the reasons why time coverage is
necessary is the so-called parallax-proper motion degeneracy. The motion of a
star on the plane of the sky is a combination of its actual motion (proper
motion) with the reflection of the Earth's orbit around the Sun (parallax). If
an object is observed for less than one year, the measured (apparent) motion
vector cannot be accurately separated into the two components, while if
observations are carried out for at least one full year, one full parallax
ellipse will be covered, and the two components will be accurately disentangled.
Another important reason is that {\em Gaia} is a complex instrument, with a
scanning law that generally allows to cover the same region of the sky multiple
times with different observations (Figure~\ref{fig:scan})\footnote{For more
details: \url{https://www.cosmos.esa.int/web/gaia/iow\_20120312}.}. Because the
accuracy in position measurements is largest along the scanning direction and
the combination of {\em Gaia}'s spin and spin axis precession provides many
observations with a different scanning orientation, it is thus possible to
reconstruct the 2D sources positions much more accurately, at the same time
keeping systematic errors under control. 

{\em Gaia} is therefore designed to observe each source on average 70 times over
the five 5 years of its nominal operation timeframe. The RVS instrument, being
designed differently, gathers about 40 different observations instead of 70, on
average. This opens up a range of time-domain astrophysical studies based on
photometry (light-curves), spectroscopy (RV curves), and astrometry (centroid
motions), obtained quasi-simultaneously in case of bright sources
\citep{eyer18}. The repeated observations allow to study periodically varying
objects such as variable stars or quasars \citep{gisella,varstrometry} or 
transient objects such as novae, supernovae, and microlensing events
\citep{lukas16}. In addition repeated observations allow to detect planets
through astrometric variability, opening a discovery window that is
complementary to other techniques \citep{perryman14}. They also allow to fully
model binary stars, which are a fundamental ingredient for understanding the
formation and evolution of stars and stellar systems \citep{eyer15}: the first
release containing non-single stars measurements will be DR3 (see
Section~\ref{sec:dr3}).

\section{Data releases and science highlights}
\label{sec:results}

Hundreds of scientists and engineers in Europe and in the world have contributed
a significant fraction of their carreers to designing, building, launching, and
operating {\em Gaia}, but also to process, analyze, validate, and publish its
data in a form that the community can use for their scientific research, for
teaching, and for outreach activities and events. Two public data releases have
taken place so far \citep{gaia1,gaia2}, and at least two more are
foreseen\footnote{Official release scenario:
\url{https://www.cosmos.esa.int/web/gaia/release}.}. Indeed, the analysis of
{\em Gaia} data in the DPAC is subdivided in 9 different Coordination Units
(CUs) that take care of various aspects such as: sowftware and database
infrastructures, data simulations, astrometric, photometric, and spectroscopic
processing, time-series analysis of variable and peculiar objects, and the like.
Closing the loops of communication, validation, and data exchanges among the CUs
smoothly is a slow and careful process of growing complexity. Therefore, each
data release presents more data products and improves on the quality of
previously released products. 

\subsection{The first {\em Gaia} data release}

The first {\em Gaia} data release, in 2016 \citep{gaia1} contained: positions
and G magnitudes of more than billion sources; the TGAS (Tycho-Gaia Astrometric
Solution) catalogue, obtained by combining Hipparchos and Gaia positions and
thus limited to G$\simeq$12 mag; and light-curves for a few thousand variable
stars at the ecliptic poles. It was thought as a demonstrational release, to
show the {\em Gaia} mission's potential and to prepare the community for the
upcoming releases. It was instead used intensively not only to test ideas and
algorithms to use in future releases, but also to obtain many new and original
scientific results in different research fields. The results were published in
more than 1000 refereed papers, that cited the {\em Gaia} first release papers
between 2016 and 2018 alone.

Many projects were carried out to test the data and methods in view of DR2. The
rotation of the Large Magellanic Cloud was studied with 29 TGAS stars
\citep{vandermarel16}; a similar study was later carried out with DR2, using 8
million stars, and discovering for the first time rotation in the Small
Magellanic Cloud \citep{gaia_gc}. The quality of TGAS parallaxes was compared
with various catalogues, and a small bias was found in the comparison with
eclipsing binaries  and previously published astrometry, of $\simeq$2.5~mas
\citep{stassun17,jao18a}, while other indicators like variable stars
\citep{gisella,casertano17} were found to be in good agreement. Finally, the
problem of converting parallaxes into meaningful distances for individual stars
was tackled using TGAS data \citep{astra16}.

The modeling of the Milky using the new DR1 and TGAS data, one of the main goals
of the {\em Gaia} mission, was carried out by various groups. I list here just
some of the most cited works: the mass distribution and halo substructure with
{\em Gaia}, RAVE, and APOGEE was studied in some detail
\citep{helmi17,bonaca17}; the dynamics of the galactic bar and the Hercules
stream were studied by \citep{monari17}; the galactic rotation was studied by
\citep{bovy17}; and a 3D study of the Orion region revealed an age gradient in
the OB-star population \citep{zari17}. {\em Gaia} DR1 and TGAS data were also
used to study Galaxy dynamics and stellar structure and evolution, by compiling
catalogues of comoving pairs and wide binaries \citep{oh17,andrews17} or very
cool stars \citep{smart17}, or by deriving fundamental relations such as the
mass-radius relation for white dwarfs \citep{tremblay17}, or finally by testing
and recalibrating various relations for pulsating variable stars
\citep{gisella}.  

Among the many scientific studies, a few received attention because they
presented new results and were covered in various {\em Gaia} press releases,
{\em Gaia} stories, or Images of the Week: \citep{koposov17} discovered two new
stellar clusters, one very close to Sirius, and thus difficult to study with
more traditional instruments\footnote{Although the discovery of the Sirius star
cluster was presented by \citet{koposov17} as entirely new, a simple ADS search
showed that the cluster was already known in the past \citep{auner80}.
Nevertheless, {\em Gaia} is the only current instrument that allows to study
such cluster in detail.}; the first supernova discovered by {\em Gaia} was
published by \citep{lukas16} as part of the science alerts program and its
ground-based follow-up network; RR~Lyrae stars were used to discover a tidal
bridge between the Large and Small Magellanic Clouds \citep{belo17}. {\em Gaia}
DR1 was also used to observe new gravitationally lensed Quasars at high redshift
\citep{lemon17,ostro18}.

\subsection{The second {\em Gaia} data release}

The second {\em Gaia} data release was issued in April 2018, and was a large
improvement over the previous one. It was the first release to include: pure
{\em Gaia} astrometry down to $G\simeq21$~mag and $G_{\rm{BP}}-G_{\rm{RP}}$
colors for more than one billion sources; mean RV measurements for more than
seven million sources; asteroid astrometry for 14\,000 known objects; and
astrophysical parameters (surface temperature and interstellar extinction) for
approximately 100 million objects. {\em Gaia} DR2 data have such exquisite
precision, that the internal structure of the data starts to be visibile in the
form of small systematic effects such as a parallax bias of 0.025~mas \citep[ten
times smaller than in DR1 TGAS,][]{lindegren18,gaia_gc}, or the G magnitude
trend of 4~millimag per magnitude \citep{evans18}. These systematic effects are
expected to decrease with each upcoming data release, as more data -- with
different characteristics -- are processed and the processing pipelines are
progressively refined.

At the time, more than 1800 refereed papers cite the second {\em Gaia} data
release paper. According to the NASA ADS, the majority of the new publications
are related to the fields of Galactic halo and disk studies (see below), but
also to solar system studies \citep{cellino07,spoto18}, exoplanet and host star
characterization \citep{kervella19}, characterization of variable or binary
objects \citep{ziegler18} and of objects with astroseismological data
\citep{berger18}. Work was done also in the field of extragalactic studies, for
example on Quasars variability \citep{varstrometry} or gravitational lensing
\citep{wertz19}. {\em Gaia} DR2 data have even been used to search for a
plausible home star for the interstellar object 'Oumuamua \citep{oumuamua},
which recently was discovered transiting in the solar
system\footnote{\url{http://www.ifa.hawaii.edu/info/press-releases/interstellar/}}.
Summarizing such a huge and diverse body of literature in a fair and complete
way is out of the scope of the present paper, so I will just present a few
highlights of galactic and stellar science from the ESA {\em Gaia} press
releases, in-depth stories, and Images of the Week.

On the Milky Way studies front, a lot of work has been done on stellar clusters,
including a full census and dynamical characterisation of open clusters
\citep[with many new discovered clusters,][]{tristan18,tristan19} and globular
clusters \citep{vasiliev19}, including the internal dynamics and substructure of
individual clusters \citep{franciosini18,bianchini18}. An impressive study on
the distribution of thousands of young clusters and OB association showed that
they lie along filamentary structures across the galaxy disk, whose orientation
changes systematically with age \citep{kounkel19}. New streams, kinematic
substructures and past accretion events have been found
\citep{helmi17,koppelman18}, from the inner galaxy to the outer halo, and
studies of dwarf galxies in the local group were carried out as well
\citep{fritz18}. Various tracers were used to further explore the galaxtic
structure, such as cepheids to trace the warp of the galactic disk
\citep{ripepi19}. 

Even relatively old research fields like stellar structure and evolution
received a boost from {\em Gaia}: given the extremely large statistics and high
precision of the of {\em Gaia} DR2 photometry, a new feature in the CMD of the
galaxy was found \citep{jao18b}, i.e., a gap along the main sequence that was
predicted theoretically but never confirmed observationally\footnote{A search
{\em a posteriori} in other survey data, such as 2MASS, indeed showed that the
gap was present, but there was not enough statistics to confirm its reality.},
caused by the transition of M stars from the fully convective to the partially
convective regime. Similarly, the sequence of white dwarfs, now very populous in
the {\em Gaia} DR2 data, showed for the first time its detailed substructure,
not only the double sequence traced by different types of white dwarfs, but also
the piling-up of white dwarfs towards the bottom of the cooling sequence, seen
in {\em Gaia} data for the first time, and likely caused by internal processes
of crystallization \citep{tremblay19}.

\subsection{Upcoming data releases}
\label{sec:dr3}

The third {\em Gaia} data release will occur in two stages in 2020 and 2021. The
early release, EDR3, will contain new astrometry and integrated photometry based
on the first 34 months of observations. The measurements uncertainties in both
astrometry and photometry will improve thanks to the increase in the number of
observations for each source, and the refinements or the addition of various
pipelines to produce the data (for example: crowding treatment, binary stars,
and extended objects). This, in turn, will increase the number of sources that
pass the quality filtering, bringing both the expected errors and the number of
sources closer to the expected end-of-mission performances. Some internal
systematics that were present in DR2, such as the parallax bias of about
0.03~mas \citep{lindegren18,gaia_gc} and the trends seen in photometry of about
4~millmags/mag or the bright blue stars systematics \citep{evans18} are expected
to improve significantly. 

The full EDR3 release, expected in 2021, will complement EDR3 with a wealth of
additional data products obtained from the same 34 months of data. DR3 will
improve incrementally on the quality of all DR2 data products (astrometry and
integrated photometry will be those of EDR3), but will also include entirely new
ones, such as: mean BP/RP spectra, non-single stars catalogues\footnote{In
particular, a full binary treatment is foreseen for all stars that show evidence
of multiplicity, including full orbital solution when possible. The
non-single-stars pipelines will be relatively simple in DR3, but as any other
processing chain in DPAC (crowding treatment, extended objects, variables, and
so on), they will improve and increase in detail and complexity from release to
release.}, results for Quasars and extended objects, object classification and
parametrization, and an additional data set, Gaia Andromeda Photometric Survey
(GAPS), consisting of the photometric time series for all sources located in a
5.5 degree field centred on the Andromeda galaxy. 

Expected presumably in 2023, DR4 will be the last release of the nominal
five-year observations period. As such, it will contain all the obtained data,
reaching the expected end-of-mission quality, and all connected data products,
such as mean and epoch data from all instruments, and results from all
pipelines. Finally, as mentioned in the introduction, the {\em Gaia} mission
extension of two more years has been finally approved and observations are
already ongoing, including a scanning law reversal period that will help in
breaking the remaining degeneracies in the astrometric solution and will bring
understanding of remaining systematic effects. A further extension of two more
years, for {\em Gaia} operations until 2022 has also been pre-approved. The
on-board fuel reserve is estimated to be sufficient to operate {\em Gaia} at
least until 2024, thus more extensions are in principle possible, and data
releases beyond DR4 are to be expected.

\section{Conclusions}
\label{sec:concl}

To achieve the required quality in astrometric measurements, {\em Gaia} is also
collecting high-quality photometry and spectroscopy, and repeatingly observing
two billion point-like sources on the sky. The resulting dataset is huge,
preriodically released and availale, to be freely explored not only by
professional astronomers, but also by amateurs, outreach professionals, and
teachers at all levels. 

Among the published papers that cite {\em Gaia}, a trend has emerged of
combining {\em Gaia} data with large existing datasets like asteroiseismic
studies, multiband photometry, or large spectroscopic surveys for the
determination of stellar abundance ratios. This trend confirms that astronomy
research is progressing towards a multi-dimensional kind of data exploration,
that can simultaneously take into account a diversity of astrophysical
properties, to gain a deeper insight. This is true both for large data samples,
where data mininig and machine learning techniques are employed to statistically
decipher the collective behaviour of astrophysical sources prior to physical
modelling, or for small data sets, where rare or new objects can be studied in
depth. Astronomers are -- necessarily -- not specialized in {\em all} of the
data analysis aspects and techniques involved, yet datasets like {\em Gaia}
allow them to move beyond their specific expertise. Each {\em Gaia} release is
accompanied by extensive documentation and various caveats that help researcher
to use the data with care and profit.

Considering all this, together with the huge data volume, involving millions and
billions of sources, {\em Gaia} is opening up a new path for astrophysics,
providing a rich dataset that is not only producing many new discovery already,
but can connect large exisiting surveys and enhance the scientfic harvest of
upcoming projects and facilities, like the James Webb Space Telescope, the
Extremely Large Telescope, or the Large Synoptic Survey Telescope.



\bibliographystyle{elsarticle-harv}
\bibliography{Gaia6D}









\end{document}